\documentclass[tradiabstract,longauth,twocolumn]{aa}  
\usepackage{graphicx}
\usepackage{txfonts}
\usepackage{hyperref}
\usepackage{xspace}
\usepackage{epstopdf, epsfig}
\usepackage[usenames]{color}
\usepackage{mathrsfs}
\usepackage{comment}

\newcommand\phcms{ph\,cm$^{-2}$\,s$^{-1}$\xspace}%

\newcommand\g{\ensuremath{\gamma}}%
\newcommand\hess{H.E.S.S.\xspace}
\newcommand\fermi{\textit{Fermi}-LAT\xspace}

\newcommand\snrG{\object{G349.7$+$0.2}\xspace}

\def\gr{$\gamma$-ray}
\def\grs{$\gamma$-rays}

\def\eg{{\it e.g.~}}

\def\ie{{\em i.e.~}}

\begin{document}

   \title{H.E.S.S. detection of TeV emission from the interaction region between the supernova remnant G349.7+0.2 and a molecular cloud}

   \author{H.E.S.S. Collaboration
\and A.~Abramowski \inst{1}
\and F.~Aharonian \inst{2,3,4}
\and F.~Ait Benkhali \inst{2}
\and A.G.~Akhperjanian \inst{5,4}
\and E.O.~Ang\"uner \inst{6}
\and M.~Backes \inst{7}
\and S.~Balenderan \inst{8}
\and A.~Balzer \inst{9}
\and A.~Barnacka \inst{10,11}
% ...
}

\institute{
Universit\"at Hamburg, Institut f\"ur Experimentalphysik, Luruper Chaussee 149, D 22761 Hamburg, Germany \and
Max-Planck-Institut f\"ur Kernphysik, P.O. Box 103980, D 69029 Heidelberg, Germany \and
Dublin Institute for Advanced Studies, 31 Fitzwilliam Place, Dublin 2, Ireland \and
National Academy of Sciences of the Republic of Armenia,  Marshall Baghramian Avenue, 24, 0019 Yerevan, Republic of Armenia  \and
Yerevan Physics Institute, 2 Alikhanian Brothers St., 375036 Yerevan, Armenia \and
Institut f\"ur Physik, Humboldt-Universit\"at zu Berlin, Newtonstr. 15, D 12489 Berlin, Germany \and
University of Namibia, Department of Physics, Private Bag 13301, Windhoek, Namibia \and
University of Durham, Department of Physics, South Road, Durham DH1 3LE, U.K. \and
GRAPPA, Anton Pannekoek Institute for Astronomy, University of Amsterdam,  Science Park 904, 1098 XH Amsterdam, The Netherlands \and
Obserwatorium Astronomiczne, Uniwersytet Jagiello{\'n}ski, ul. Orla 171, 30-244 Krak{\'o}w, Poland \and
now at Harvard-Smithsonian Center for Astrophysics,  60 Garden St, MS-20, Cambridge, MA 02138, USA
% ...
}

   \authorrunning{The H.E.S.S.~collaboration}
   \titlerunning{Detection of G\,349.7$+$0.2 with H.E.S.S.}
   
    \offprints{D.~Fernandez, C.~Trichard \\
     \email{\href{mailto: diane.fernandez@lupm.univ-montp2.fr}{diane.fernandez@lupm.univ-montp2.fr}, \\
    		\href{mailto: cyril.trichard@lapp.in2p3.fr}{cyril.trichard@lapp.in2p3.fr}} 
     }

  \date{Accepted November 24, 2014}

\abstract{
\snrG~is a young Galactic supernova remnant (SNR) located at the distance of 11.5 kpc and observed across the entire electromagnetic spectrum from radio to high energy (HE; 0.1 GeV < E < 100 GeV) $\gamma$-rays. Radio and infrared observations indicate that the remnant is interacting with a molecular cloud. In this paper, the detection of very high energy (VHE, E > 100 GeV) \gr~emission coincident with this SNR with the High Energy Stereoscopic System (H.E.S.S.) is reported. This makes it one of the farthest Galactic SNR ever detected in this domain. An integral flux $F(E>400\,\mathrm{GeV}) = (6.5 \,\pm 1.1_\mathrm{stat} \,\pm 1.3_\mathrm{syst}) \times 10^{-13}\,$ \phcms corresponding to $\sim$0.7\% of that of the Crab Nebula and to a luminosity of $\sim10^{34}$~$\rm erg\,s^{-1}$ above the same energy threshold, and a steep photon index $\Gamma_{\rm VHE} = 2.8 \,\pm 0.27_{\mathrm{stat}} \,\pm 0.20_{\mathrm{syst}}$ are measured. The analysis of more than 5~yr of \fermi~data towards this source shows a power-law like spectrum with a best-fit photon index $\Gamma_{\rm HE} = 2.2 \,\pm 0.04_{\rm stat}\,{}^{+0.13}_{-0.31}{}_{\rm sys}$.
The combined \gr~spectrum of \snrG~can be described by either a broken power-law (BPL) or a power-law with exponential (or sub-exponential) cutoff (PLC). In the former case, the photon break energy is found at $E_{\rm br, \gamma} = 55^{+70}_{-30}$ GeV, slightly higher than what is usually observed in the HE/VHE \gr~emitting middle-aged SNRs known to be interacting with molecular clouds. In the latter case, the exponential (respectively sub-exponential) cutoff energy is measured at $E_{\rm cut, \gamma} = 1.4^{+1.6}_{-0.55}$~(respectively $0.35^{+0.75}_{-0.21}$)~TeV. A pion-decay process resulting from the interaction of the accelerated protons and nuclei with the dense surrounding medium is clearly the preferred scenario to explain the \gr~emission. The BPL with a spectral steepening of 0.5-1 and the PLC provide equally good fits to the data. The product of the average gas density and the total energy content of accelerated protons and nuclei amounts to $n_{\rm H}\,W_{\rm p} \sim 5 \times 10^{51}$~erg cm$^{-3}$.}


      \maketitle

\section{Introduction}

The question of the origin of Galactic Cosmic Rays (CRs) dates back one century. In the 1930s, \citet{1934PNAS...20..259B}  proposed supernovae (SNe) as probable sources of Galactic CRs. According to the diffusive shock acceleration (DSA) theory \citep[\eg][]{1978MNRAS.182..147B, 1978MNRAS.182..443B} particles are accelerated at the supernova remnant (SNR) shock waves. The spectrum of the accelerated particles follows a power-law shape with exponential cutoffs and spectral indices of $p \sim 2$, compatible with radio measurements.  Such spectra have also been observed in $\gamma$-rays from several isolated SNRs \citep[\eg][]{2007A&A...464..235A}. However, recent \fermi~observations of SNRs interacting with molecular clouds (MC) have revealed spectral breaks above a few GeV \citep{fermi_W51C_2009, fermi_w28_2010, 2010Sci...327.1103A, fermi_IC443_2013, fermi_W49B_2010, 2013ApJ...774...36C}. 

\snrG~is a bright Galactic SNR with a small angular size of $\sim 2.5'\times 2'$ \citep{2009BASI...37...45G} and a roughly circular morphology similar in radio \citep{1985Natur.313..113S} and X-rays \citep{2002ApJ...580..904S,2005ApJ...618..733L}. The brightness enhancement seen towards the southwest of the SNR suggests that \snrG~is expanding into a density gradient caused by a HI cloud. Indeed, the coincidence of \snrG~with a dense MC \citep{2004A&A...426..201D} and the detection of five OH (1720 MHz) masers towards the centre of the SNR \citep{1996AJ....111.1651F} and of line emissions from several molecular transitions \citep{2000ApJ...545..874R,lazendic2010} provide evidence in support of an interaction between the SNR and the MC. These masers and molecular line emissions are measured at similar velocities which, together with HI absorption measurements, originally placed the SNR at a distance of $\sim$ 22.4 kpc. 
\cite{2014arXiv1401.3354T} have revised the kinematic distance to $\sim$11.5~kpc based on updated knowledge of the kinematics in the inner Galaxy \citep{2008ApJ...683L.143D,2008A&A...489..115R} together with high-resolution 21~cm~HI  (from the Southern Galactic Plane Survey, SGPS; \cite{2012ApJS..199...12M}) and CO data \citep{2000ApJ...545..874R}. Thus, \snrG~is located at the near edge of the Far 3 kpc Arm rather than on the far side of the Galaxy. This distance estimate was confirmed by \cite{yasumi_2014}.
At the revised distance, the SNR radius and age are $\sim$ 3.8 pc and $\sim$ 1800 yr, respectively. The overall X-ray emission of \snrG~is best fit with two thermal components from the shocked SN ejecta and circumstellar material, and results in a blast wave velocity estimate of $\sim$ 700$-$900 km s$^{-1}$ \citep{2002ApJ...580..904S,2005ApJ...618..733L}. In the high-energy (HE; 0.1 GeV $<$ E $<$ 100 GeV) \gr~domain, \citet{2010ApJ...717..372C} discovered an unresolved \g-ray source coincident with \snrG~based on \fermi~observations, designated as 2FGL J1718.1-3725 in the two-year \fermi catalog \citep{2012ApJS..199...31N}. The spectrum was best fit with a simple power-law with $\Gamma_{\rm HE}$ = 2.1 $\pm$ 0.1, and the addition of an exponential cutoff was found to only marginally improve the fit.

In this paper the detection of very high energy (VHE, E $>$ 100 GeV) $\gamma$-ray emission coincident with this SNR in observations with the High Energy Stereoscopic System (H.E.S.S.) experiment is reported. \hess~observations and data analysis results are presented in Sect.~\ref{sec:Ana}, together with the analysis of more than 5 yr of \fermi~data towards \snrG. Based on all the available multi-wavelength data, the SNR-MC scenario to account for the broadband spectral energy distribution of \snrG~is discussed in Sect.~\ref{sec-discu}, in the light of recent theoretical works aimed at explaining the \gr~spectrum of such interacting SNRs.

\section{Analysis}
\label{sec:Ana}
 
\subsection{\hess observations}
\label{sec-hess}
\hess~is an array of five imaging atmospheric Cherenkov telescopes (IACTs) located in the Khomas Highland of Namibia at an altitude of 1800~m above sea level  \citep{2006A&A...457..899A}. The fifth telescope (28-m diameter) started operation in September 2012. All H.E.S.S. data used in this paper have been taken with the four-telescope array, which detects $\gamma$-rays above an energy threshold of $\sim$ 100\,GeV and covers a field of view of 5\degr diameter. The primary particle direction and energy are reconstructed with event-by-event resolutions of $\sim$ 0.1\degr~and $\sim$ 15\%, respectively.

The data set for the source analysis includes observations taken from 2004 to 2012 and is summarized in Table~\ref{tab:DatasetTab}. Two data sets are made up of Galactic Scan runs from previous Galactic Plane surveys in 2004 and 2008 \citep{2013arXiv1307.4690C}. A set of 24 dedicated runs were taken using the so-called \emph{wobble} mode for which the source is alternatively offset from the pointing direction by a small distance varying from 0.40\degr\ to 0.75\degr. This method allows for the evaluation of the signal and the background from the same observation. A fourth data set is composed of \emph{wobble} runs dedicated to the observation of other nearby sources, in particular {\object{RX\,J1713.7$-$3946}\xspace} \citep{2007A&A...464..235A} located at $\sim$ 2.5\degr\ from \snrG. The total data set comprises 113 hours of observations (live time) after applying quality cuts. 
\begin{table*}
  \caption{Details of the data set for the analysis of \snrG}
  \label{tab:DatasetTab}  
  \centering
  \begin{tabular}{ccccccc}
    \hline\hline
    Data set 		    		& Date      		    & Live Time (hours) & Number of runs 		 &  Offset  (Mean offset) (\degr ) \\
    \hline
    Galactic Scan 1            		& 05-07/2004             & 7.2            		    & 17      			&  0.6$-$2.3  \, (1.6) \\
    Galactic Scan 2         		& 05-06/2008             & 11.8         		    & 28       		&  0.7$-$1.5  \, (0.9) \\
    \snrG~\emph{wobble} runs    & 04-09/2010              &10.5         		    & 24      			&  0.5$-$0.7  \, (0.5)  \\
    Other sources            		& 04/2004-09/2012   & 83.5         		    &194     			&  0.8$-$2.3 \, (1.9)  \\
    \hline
  \end{tabular}
\end{table*}

Data have been analysed with the Model Analysis as described in \citet{2009APh....32..231D} and using \textit{Standard cuts}. The analysis has been cross-checked with an independent data calibration chain and multivariate analysis method \citep{2009APh....31..383O}. 
The extraction region is defined as a circular region of radius $\theta = 0.1\degr$ centred on the nominal position of the X-ray source \snrG~from the Chandra Supernova Remnant Catalog\footnote{http://hea-www.cfa.harvard.edu/ChandraSNR/G349.7+00.2/}: $\alpha_\mathrm{J2000}=17^\mathrm{h} 17^\mathrm{m} 59\fs6$, $\delta_\mathrm{J2000}=-37\degr 26\arcmin 21\farcs3$.
After background subtraction with the reflected background method \citep{Berge_2007}, an excess of 163 VHE \g-rays is detected within the analysis region, which corresponds to a significance level of 6.6~$\sigma$ 
according to Eq. 17 from \cite{1983ApJ...272..317L}. 
Given the existence of GeV emission and a source extent of $\sim2.5'\times2'$ (much smaller than the \hess~PSF), an unresolved VHE \gr~signal was expected and only one source position and extent have been tested.
The excess is point-like within the \hess point spread function (PSF) uncertainties and the best fit position of the VHE emission within the extraction region is found to be $\alpha_\mathrm{J2000}=17^\mathrm{h} 17^\mathrm{m} 57\fs8  \pm 2\fs0_\mathrm{stat} \pm 1\fs3_\mathrm{syst}$, $\delta_\mathrm{J2000}=-37\degr 26\arcmin 39\farcs6 \pm 24\farcs0 _\mathrm{stat} \pm 20\farcs0_\mathrm{syst}$, compatible with the X-ray position of \snrG. An upper limit on the source extent of 0.04\degr\ (95\% confidence level, CL), larger than the SNR size seen in radio and X-rays, is obtained based on the log-likelihood method profile. 

\begin{figure}[h]
\centering
\includegraphics[width=\columnwidth]{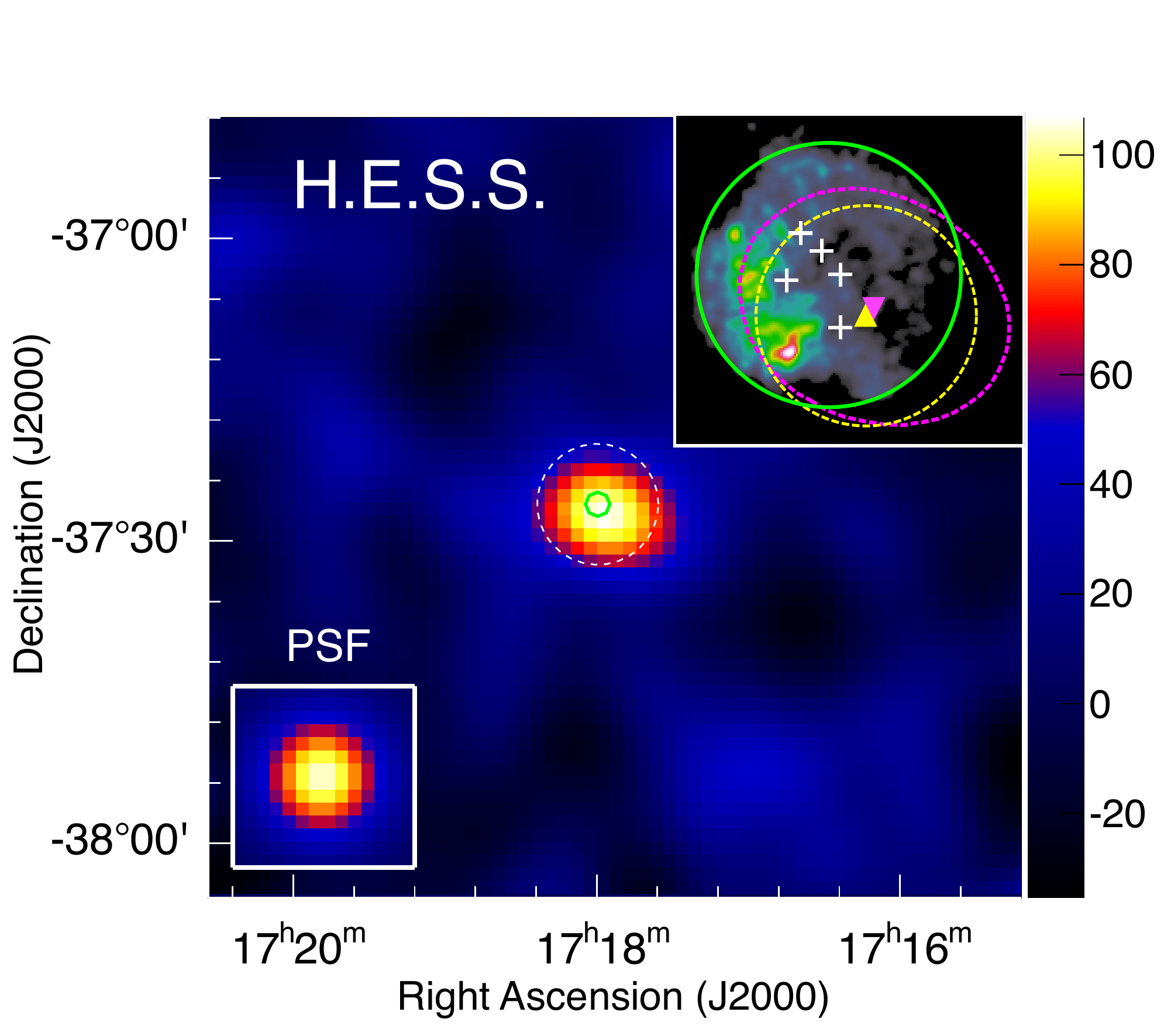}
\caption{\hess \g-ray excess map of \snrG. The image is smoothed with a gaussian with a width of $0.06\degr$ corresponding to the PSF of the analysis (shown in the bottom left inset). The color scale represents the excess counts per surface area of $\pi (0.06\degr)^{2}$. Pixels within this area are correlated. The solid green and dashed white circles denote the \snrG~radio shell and the \hess ON region, respectively. The upper right inset represents the {\it Chandra} image of \snrG~with the five OH (1720 MHz) masers (white crosses) delineating the associated MC as found by \cite{2004A&A...426..201D} . The best fit position together with its 2$\sigma$ CL contours of the TeV emission are marked with a magenta inverted triangle and magenta dashed contours, respectively. The {\it Fermi}-LAT best fit position and its 2$\sigma$ CL contours are shown as a yellow triangle and a yellow dashed contour. The green circle denoting the \snrG~extent is reproduced in the inset for scaling.} 
\label{fig:map}
\end{figure}

Figure \ref{fig:map} shows the excess count image smoothed with a Gaussian of width 0.06\degr~which corresponds to the 68\% containment radius of the \hess PSF for this analysis. The SNR \snrG~and the \hess analysis region are indicated by solid green and white dashed circles, respectively. As seen in the inset image, the 2 $\sigma$ error contours of the \hess~best fit position show that the position of the VHE source is compatible with the whole SNR as observed with {\it Chandra} as well as with the five OH (1720 MHz) masers.

\begin{figure}[h]
\centering
\includegraphics[width=0.4\textwidth]{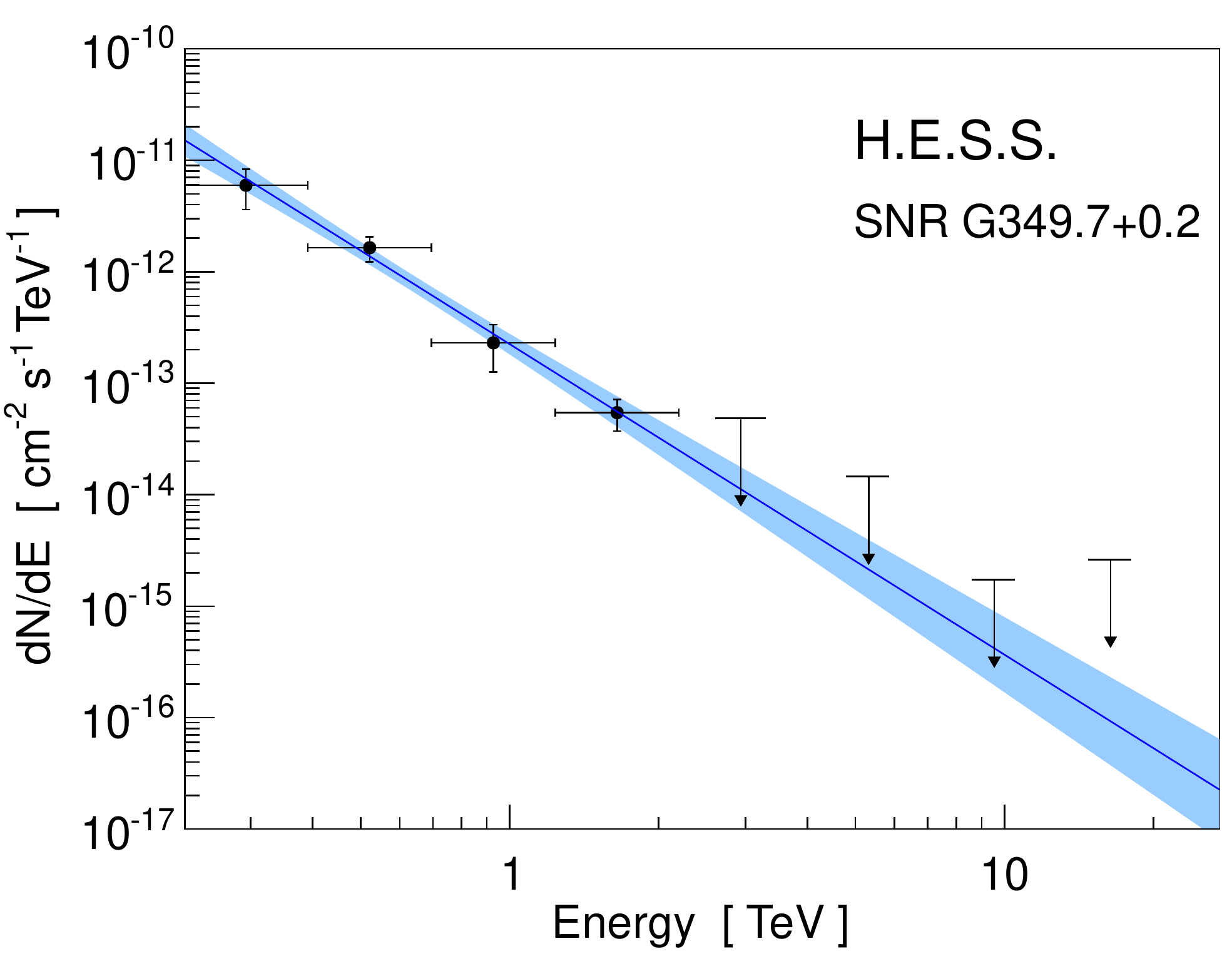}
\caption{\hess forward folded spectrum of \snrG. The blue line is the best fit of a power-law to the data as a function of the energy (unfolded from the H.E.S.S. response functions). The blue bowtie is the uncertainty of the fit given at 68\% CL. Upper limits are given at 99\% CL.} 
\label{fig:spec}
\end{figure}
  
The energy spectrum of the VHE emission coincident with \snrG~is extracted above 220 GeV and fitted using the forward folding technique described in \citet{2001A&A...374..895P}. The resolution-unfolded spectrum is shown on Figure \ref{fig:spec}. 
The spectrum is well described by a power-law defined as $d\Phi/dE \propto E^{-\Gamma}$ with a photon index of $\Gamma_{\rm VHE} = 2.8 \pm0.27_{\mathrm{stat}} \pm0.20_{\mathrm{syst}}$ ($\chi^2/n_\mathrm{dof} = 54.1/56$). The integrated photon flux above 400 GeV is $F(E>400\,\mathrm{GeV}) = (6.5 \pm1.1_\mathrm{stat} \pm1.3_\mathrm{syst}) \times 10^{-13}$\,\phcms which corresponds to 0.7\%  of the Crab Nebula flux \citep{2006A&A...457..899A} and to a luminosity of $\sim 10^{34}$~$\rm erg\,s^{-1}$ above the same energy threshold. Spectral models of a curved power-law and a power-law with exponential cutoff do not improve the fit of the spectrum significantly.

\subsection{\fermi~observations} 
\label{sec-fermi}

The LAT detector is the main instrument on board the {\it Fermi Gamma-Ray Space Telescope (Fermi)}. It consists of a pair-conversion imaging telescope detecting \gr~photons in the energy range between 20 MeV and $\gtrsim$ 300 GeV, as described by \citet{atwood2009}. 
The LAT has a effective area of $\sim$ 8000 cm$^{2}$ on-axis above 1 GeV, a field of view of $\sim$ 2.4 sr and an angular resolution of $\sim$ 0.6\degr\ (68\% containment radius) at 1 GeV for events converting in the front section of the tracker\footnote{More information about the performance of the LAT can be found at the FSSC: \texttt{http://fermi.gsfc.nasa.gov/ssc}}.

A GeV \gr\ excess associated with \snrG~was first reported by~\citet{2010ApJ...717..372C} using 1 year of \fermi~data. Since this discovery, several improvements have been made both in the instrument response functions (IRFs) and in the data analysis software. The following analysis was performed using 5.25 yr of data collected from 2008 August 4 to 2013 November 6. The latest version of the publicly available Fermi Science Tools\footnote{\texttt{\tiny http://fermi.gsfc.nasa.gov/ssc/data/analysis/documentation/Cicerone/}} \emph{(v9r32p5)} was used, with the \emph{P7REP\_SOURCE\_V15} IRFs and the user package \texttt{enrico}~\citep{2013arXiv1307.4534S}. Events at normal incidence ($\cos(\theta)>0.975$), with zenith angles smaller than 100\degr, and flagged as \emph{source class events} were selected to perform a binned likelihood analysis. A region of 10\degr~around the position of \snrG~was analysed. All sources from the \fermi~two-year source catalog ~\citep{2012ApJS..199...31N} within 12\degr~around the target were added. The ones closer than 8\degr~(i.e.~the 95\% containment radius of the LAT PSF for front- and back-converted events at 200 MeV), and with a significance larger than 3, were modeled simultaneously (with fixed positions). Additionally, the Galactic and extra-galactic diffuse models were used with their respective normalization treated as free parameter. The likelihood analysis was performed with the \texttt{gtlike} tool. To determine the significance of the signal, the test-statistic (TS) method was used: $TS=-2 \ln(\frac{L_{null}}{L_{ps}})$, where $L_{ps}$ and $L_{null}$ are the maximum likelihood value for a model with and without an additional source, respectively.

For the spatial analysis, \gr~events with 5~GeV$\,<\!E_{\gamma}\!< $ 300~GeV were selected. Such a selection provides a good instrument PSF and both low source confusion and background level from the Galactic \gr~diffuse emission. The TS in this energy band is $\sim$102. The source position, determined by \emph{gtfindsrc} tool, is $\alpha_\mathrm{J2000}=17^\mathrm{h} 17^\mathrm{m} 58\fs0 \, \pm 3\fs5_\mathrm{stat} \, \pm 4\fs0_\mathrm{syst}$, $\delta_\mathrm{J2000}=-37\degr 26\arcmin 42\farcs0 \,\pm 54\farcs0_\mathrm{stat} \,\pm 60\farcs0_\mathrm{syst}$. The systematic errors are estimated according to the 2FGL catalog~\citep{2012ApJS..199...31N}. The \fermi source is compatible with the \hess\ position at less than 1$\sigma$ and is consistent with the radio shell of \snrG, as shown on Figure \ref{fig:map}.  The \fermi~residual count map shows no evidence of a significant source extension after point like source subtraction.

For the spectral analysis, \emph{front-} and \emph{back-}converted events in the 0.2-300 GeV energy range were selected. The lower bound was chosen in order to reduce both the systematic uncertainties on the \fermi~PSF and acceptance, and the level of Galactic \gr~diffuse emission. In the following the effect of the underlying Galactic diffuse emission on the source flux was estimated by varying artificially the normalization of the Galactic background model by $\pm$6\% from the best-fit value. The systematic uncertainties related to the IRFs are not estimated here as they are usually smaller than those arising from the Galactic diffuse emission in the $\sim$0.1$-$50~GeV energy range.
A point-like power-law model at the position of the SNR was used and the best fit parameters were determined by the likelihood method. The source TS value from this analysis is  $\sim$201 and the best fit photon index 
$\Gamma_{\rm HE} = 2.2\,\pm0.04_\mathrm{stat} \,{}^{+0.13}_{-0.31}{}_{\rm sys}$. The energy flux at the decorrelation energy $E_0$ = 3.8~GeV is $F_{E_0}= (6.1 \,\pm 0.43_{\rm stat} \,{}^{+3.1}_{-2.7}{}_{\rm sys}) \times 10^{-12}\,  \rm erg\,cm^{-2}\,s^{-1}$. The flux in the full energy range is $F_{0.2-300\, \rm GeV}= (2.8 \,\pm 0.32_{\rm stat} \,{}^{+1.4}_{-1.2}{}_{\rm sys}) \times 10^{-8}\, \rm ph\,cm^{-2}\,s^{-1}$. Analyses assuming log-parabola and smoothed broken power-law spectrum models were performed in the same energy range. No improvement of the fit was found indicating that there is no significant deviation from a pure spectral power-law.
A binned spectral analysis was also performed following the same method described above in each energy bin. The resulting spectrum is shown in Figure \ref{fig:sed} and discussed in Sect. \ref{sec-discu}. A 99\% CL upper limit was calculated in the bins with a signal significance lower than 3 $\sigma$. The statistical uncertainties are given at 1 $\sigma$.

\subsection{Combined analysis} 
\label{sec-comb}

The $\gamma$-ray source detected with \fermi~and \hess~towards \snrG~shows that the object is a luminous Galactic SNR, with luminosities in the 0.2$-$300 GeV energy range and above 400 GeV of $L_{\rm HE} \sim 3 \times 10^{35}\, \rm erg\,s^{-1}$ and $L_{\rm VHE} \sim~10^{34}\, \rm erg\,s^{-1}$, respectively, assuming a distance of 11.5~kpc. 
The VHE spectrum from \snrG~is well fitted with a steep power-law shape with photon index $\Gamma_{\rm VHE} = 2.8 \,\pm 0.27_{\mathrm{stat}} \,\pm 0.20_{\mathrm{syst}}$, which represents a steepening from the one measured at HE by \fermi ($\Gamma_{\rm HE} = 2.2 \,\pm 0.04_{\mathrm{stat}}\,{}^{+0.13}_{-0.31}{}_{\rm sys}$) of $\Delta\Gamma=0.60\,\pm 0.27_{\mathrm{stat}} \,{}^{+0.23}_{-0.37}{}_{\rm sys}$. The position of the spectral break is estimated through a likelihood ratio test statistic \citep{Rolke_2005} applied to the \hess~and \fermi~data, taking both statistical and systematic uncertainties into account:
\begin{equation}
 	   \Lambda (E_{\rm br0})  = \frac{\underset{\theta} \sup_{}\, \mathcal L(E_{\rm br0}, \theta)} {\underset{E_{\rm br},\theta} \sup_{}\, \mathcal L(E_{\rm br}, \theta)},
\end{equation}
where $E_{\rm br0}$ is the tested hypothesis. The supremum in the denominator is determined over the full parameter space.
The spectral indices and the normalization of the photon spectrum are considered as nuisance parameters represented by the $\theta$ variable. They are set free. The minimum of the likelihood ratio is reached at the photon energy $E_{\rm br, \gamma}$=\,55 GeV, and the 68\% confidence interval is $[ 25; 125 ]$ GeV. The \gr~spectral steepening is thus precisely at the transition between the \fermi and \hess~domains. A cutoff $e^{-(E/E_{\rm cut})^{\beta}}$, where $E_{\rm cut}$ is the cutoff energy and $\beta$ defines the spectral shape in the cutoff region could also accommodate the steep and faint VHE spectrum at these intermediate energies. Following the same method as for the broken power-law \gr~spectrum, the spectral turnover is found to be at $E_{\rm cut, \gamma} =  1.4^{+1.6}_{-0.55}$~(respectively $0.35^{+0.75}_{-0.21}$)~TeV assuming a power-law photon spectrum with an exponential (respectively sub-exponential, $\beta$=0.5) cutoff.
The shape of the cutoff in the photon spectra with respect to that in the particle spectrum depends on the emission process, and exponential cutoffs in the particle spectrum typically result in sub-exponential cutoffs in the photon spectrum for pion decay \citep{2006PhRvD..74c4018K} and inverse Compton emission \citep{2012ApJ...753..176L}.
A power-law particle spectrum is predicted by DSA\footnote{Non-linear acceleration effects in CR modified shocks may even give rise to slightly concave spectra. An even more pronounced concave shape, steep at HE and hard at VHE \grs \,\citep{gabici2009} or a parabolic shape peaking at VHE \citep{2011ApJ...731...87E}, may occur in the case of a MC illuminated by CRs escaping from a nearby SNR.}, and a cutoff is generally formed. Such a spectrum can be interpreted as the emission from accelerated particles at the SNR shock, the cutoff being due to either escape of the highest energy particles or limitation of the acceleration because of the SNR age or radiative losses (for leptons) \citep{2007A&A...464..235A}.
On the other hand, \gr~broken power-law spectra with $E_{\rm br, \gamma}\sim$1$-$20~GeV have been observed in several SNRs known to be interacting with MCs \citep[see][and references therein]{jiang2010}, such as W28 \citep{2008A&A...481..401A,fermi_w28_2010}, W51C \citep{fermi_W51C_2009,2012A&A...541A..13A}, W49B \citep{fermi_W49B_2010,2011arXiv1104.5003B}, IC443 \citep{2009ApJ...698L.133A,fermi_IC443_2013} or W41 \citep{2006ApJ...636..777A,2013ApJ...774...36C}.
The CR spectral shape (broken power-law and exponential cutoff power-law) underlying this \g-ray~spectrum will be investigated in the following sections in view of the $\gamma$-ray emission scenarios.

\section{Discussion}
\label{sec-discu}

\subsection{Multi-wavelength considerations}
\label{gnrl}

In order to address the question of the origin of the \gr~emission from \snrG, the published radio and X-ray data from the SNR have been assembled. Radio flux densities are provided by \citet{2009BASI...37...45G} ($F_{\nu}(1\mathrm{GHz}) = 20\rm\,Jy$) and \citet{1976MNRAS.174..267C} ($F_{\nu}(408\,\mathrm{MHz}) = 31\rm\,Jy$, $F_{\nu}(5\,\mathrm{GHz}) = 9.1\rm\,Jy$). X-ray observations of \snrG~with {\it Chandra} have revealed the thermal nature of the SNR emission, from both the ejecta and shocked circumstellar medium~\citep{2005ApJ...618..733L}. A power-law (non-thermal) component was estimated to contribute to less than $2.6 \%$ (at $3 \sigma$ CL) of the total flux in the $0.5-10$~keV range for any photon index between 1.5 and 3. This translates into a flux upper limit of $1.7\times10^{-11} \rm erg\,cm^{-2}\,s^{-1}$. 
A post-shock Hydrogen density of $\mathbf{\sim 7 \rm\,cm^{-3}}$, leading to an ISM density of $\mathbf{\sim 1.7 \rm\,cm^{-3}}$ under the assumption of a strong shock was derived from the soft component of the SNR thermal X-ray spectrum \citep{2005ApJ...618..733L}.
From $\rm^{12}CO$ observations, \citet{2004A&A...426..201D} reported that \snrG~is associated with a MC, whose total mass and average density are estimated to be of $M_{\rm MC}\sim~5 \times 10^{3}\,\rm M_{\odot}$ and $n_{\rm MC}\sim 2 \times 10^{4} \,\rm cm^{-3}$ at 11.5 kpc, respectively. 
Another density estimate comes from the presence of 5 OH (1720 MHz) masers \citep{1996AJ....111.1651F} and strong $\rm H_{2}$ lines \citep{2009ApJ...694.1266H} towards the centre of the remnant, both tracers originating from shocked molecular region of very high density ($n \sim 10^{4..6}\rm\,cm^{-3}$). As discussed by \cite{2005ApJ...618..733L} and \citet{2010ApJ...717..372C}, these differences in density estimates indicate that the SNR is expanding in an inhomogeneous, likely clumpy, medium.

\subsection{SNR shell emission}
\label{sec:shell}

To quantify the total amount of energy required to explain the \gr~spectrum, a simple time-independent one-zone model of accelerated particles and their
associated broadband emission spectra is compared to the multi-wavelength (radio and X-ray) data described in the previous section. 
A power-law with exponential cutoff model for the CRs spectrum is adopted:  $dN/dE \propto E^{-p} \exp(- E/E_{\rm cut})$.
Typical values for the SN explosion energy and for the fraction that goes into CR acceleration are assumed: $E_{\rm SN}= 10^{51} \rm\,erg$ and $\epsilon_{\rm CR} \sim 0.1$ (\ie~$W_{\rm p}+W_{\rm e} = \epsilon_{\rm CR}E_{\rm SN}$, where W$_{\rm p,e}$ are the total amount of explosion energy going into protons and electrons acceleration, respectively). 
Photon spectra from non-thermal Bremsstrahlung (NBr), Inverse Compton (IC) and proton-proton (p-p, followed by $\pi^0 \rightarrow 2 \gamma$) processes are computed according to \cite{1970RvMP...42..237B,1999ApJ...513..311B,Kafexhiu_2014} \citep[the hadronic emission is multiplied by the factor $\sim1.5$ to take into account nuclei heavier than Hydrogen,][]{1986A&A...157..223D}.

A NBr-dominated scenario requires an electron to proton ratio $K_{\rm ep}\geq0.2$, which is much higher than the values expected from CR abundances and from the modeling of the broadband emission from several SNRs, which lie in the $\sim10^{-2}-10^{-3}$ range \citep{2008JCAP...01..018K}. The IC-dominated scenario requires a spectral shape much harder than the one observed at GeV energies. Moreover, values of both the energy content in radiating electrons ($W_{\rm e} \sim 8\times10^{50}$\,(d/11.5 kpc)$^{2}$~erg) and the magnetic field (B~$\leq 4\,\mu$G) for IC on CMB are unrealistic.  
The optical interstellar radiation field from \cite{2008ApJ...682..400P} has a negligible effect on the IC emission in this region of the Galaxy, while the previous parameters change to $W_{\rm e} \sim 10^{50}$\,(d/11.5 kpc)$^{2}$~erg and B~$\leq 8\,\mu$G, when accounting for the infrared interstellar radiation fields. However the energy density of the different photon fields from \cite{2008ApJ...682..400P} is known to be subject to large uncertainties at small scales.
Because of the large electron to proton ratio and the low magnetic field required in NBr- and IC-dominated scenarios, the leptonic origin of the \gr~emission is disfavored. 
\begin{figure}[!htb]
\centering
\includegraphics[width=\columnwidth]{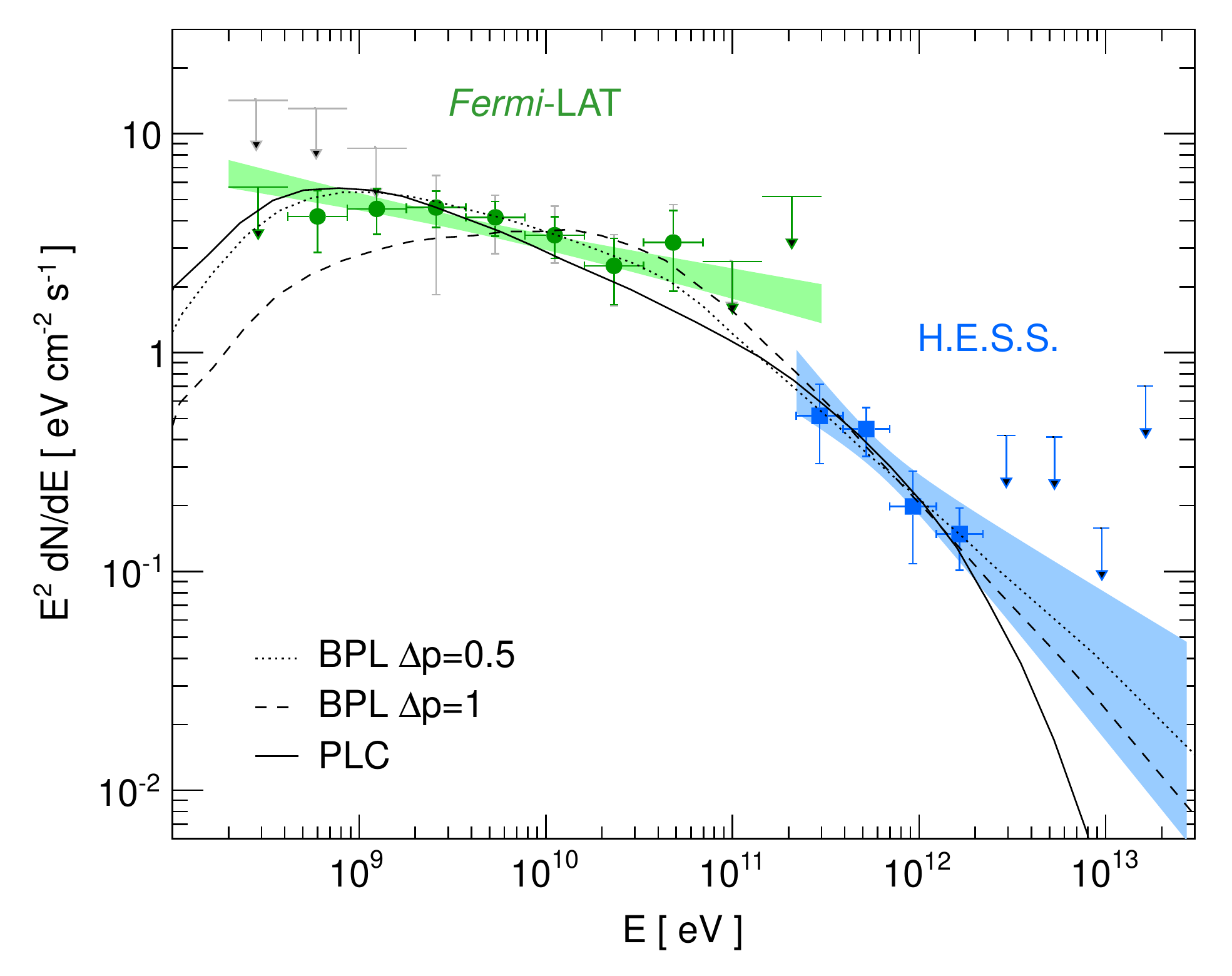}
\caption{\gr~spectrum of \snrG. The \hess (blue) and \fermi (green) spectra are shown with their 68\% CL bowtie. For the \fermi~spectral points the statistical errors are marked green while the statistical errors including the systematic errors are grey. The H.E.S.S. points are given with their statistical errors only. The $\pi^0$-decay emission spectra obtained with the best fit proton distributions are shown as dotted and dashed lines for the broken power-law (BPL) distributions with steepening $\Delta p$=0.5 and 1, respectively, and as a solid line for the power-law with exponential cutoff (PLC) distribution.} 
\label{fig:sed}
\end{figure}

The $\pi^0$ decay dominated scenario leads to a product of the average gas density and the total energy content of accelerated hadrons of $n_{\rm H}$W$_{\rm p}$~$\sim~5 \times 10^{51}$\,(d/11.5 kpc)$^{2}$~erg cm$^{-3}$, similar to what has been derived for W51C \citep{fermi_W51C_2009} and W49B \citep{fermi_W49B_2010}.
To constrain the parameters of the primary proton  distribution, its resulting photon spectrum from p-p interactions \citep[computed using the parametrization of][]{Kafexhiu_2014} is compared to the observed photon spectrum through a Markov chain Monte   Carlo (MCMC) fitting procedure using the \texttt{gammafit} package\footnote{\label{gammafitLink}\hyperref[https://github.com/zblz/gammafit]{https://github.com/zblz/gammafit}}. 
The best-fit proton spectral parameters are a spectral index $p=2.4^{+0.12}_{-0.14}$ and a cutoff energy of $E_{\rm cut} = 6.8^{+10}_{-3.4}$~TeV. With an ISM density of $\mathbf{\sim1.7 \,\rm cm^{-3}}$, as discussed in the previous section, the $\pi^0$ decay scenario would require a too large energy content in the accelerated protons and nuclei of $\sim 3 \times 10^{51}$~erg. Thus the \gr~emission coincident with \snrG~clearly can not arise from the whole SNR shell assumed to evolve in an homogeneous $\mathbf{\sim1.7 \,\rm cm^{-3}}$ ISM, but rather from the region of the SNR-MC interaction. The $\pi^0$-decay emission spectrum obtained with the best fit proton distribution is shown on Figure \ref{fig:sed}. 
In the standard modelings of gamma-ray emission from MC illuminated by CRs from a nearby, non-interacting, source, the VHE emission from these escaping CRs is expected to be harder than the HE emission from particles still confined in the source \citep{gabici2009, 2011ApJ...731...87E}. This is opposite to what is observed here from G349.7+0.2. Together with the CR energetics constraints, another scenario, in which the particular interaction region between the blast wave and the cloud at the origin of the HE/VHE emission, must be investigated.

\subsection{SNR-MC interaction scenario}
\label{origin}
 
As mentioned in Sect. \ref{sec-comb}, spectral breaks at $\sim$1$-$20 GeV have recently been observed in several interacting SNRs. These spectral features are not a priori predicted by the DSA theory and several theoretical scenarios have been put forward in order to explain \gr~spectral breaks. They can be due to either acceleration effects on particles residing within the interacting SNR \citep{inoue2010,uchiyama2010,malkov2011,tang2011,malkov2012} or diffusion of particles escaping from the SNR shock and diffusing in the MC \citep{li2010,ohira2011,li2012,aharonian1996}.
In particular ion-neutral collisions occurring when fast shocks interact with partially ionized material can lead to Alfv\'en wave damping \citep{drury1996,ptuskin2003,ptuskin2005} and hence, to the reduction of the confinement of the highest energy particles which escape the system. As shown by \citet{malkov2011} in the case of W44, and recently generalized by \citet{malkov2012} (see their Equation 4), a break naturally occurs at a few GeV, above which the particle spectrum steepens by one power $\Delta p$~=~1. 
\citet{ohira2011} have reinvestigated the distribution of CRs escaping from a SNR assumed to be of finite size, based on the escape-limited model of CR acceleration described in \citet{ohira2010}. In this model, once the forward shock approaches the MC modeled as a shell surrounding the SNR (more precisely, when the distance between the shock front and the MC inner radius equals the diffusion length of the escaping CRs) all particles are expected to escape from the SNR because of wave damping. Besides the breaks arising from the finiteness of the source and emission regions, another break, interpreted as the maximum particle energy in the SNR when it encounters the MC, is found. Both scenarios could reproduce the \gr~spectrum from \snrG, though at the expense of several free parameters related to the diffusion and the MC properties. 

The scenarios cited above assume that the \gr~emission arises from hadronic interactions of accelerated protons and nuclei with the surrounding dense medium. 
To constrain the spectral shape of the accelerated particles within these scenarios, the same MCMC method as described in Section \ref{sec:shell} was employed assuming a broken power-law for the proton spectrum. Two values were considered for the spectral steepening above the break energy: $\Delta p\,=\,1$ as predicted by \cite{malkov2011} and $\Delta p\,=\,0.5$ as the spectral steepening appears to be lower than 1 in some \gr~emitting SNR-MC systems \citep[e.g.~W28, IC443][]{fermi_w28_2010, fermi_IC443_2013}. The best-fit parameters are a HE spectral index of $p_1=2.0^{+0.40}_{-0.23}$ (respectively $p_1=2.3^{+0.15}_{-0.13}\,$) and a break energy $E_{\rm br}= 0.26^{+1.2}_{-0.22}$ (respectively $0.25^{+0.75}_{-0.20}\,$)~TeV for a steepening $\Delta p\,=\,1$ (respectively $\Delta p\,=\,0.5$). The resulting \gr~spectra are shown in Figure \ref{fig:sed}. 

One can compare the Bayesian Information Criterion $BIC = - 2 \log(L) + k \times \log(N)$, where $k$ is the number of free parameters in the model, and $N$ the number of observations, obtained for the two broken power-laws and the power-law with exponential cutoff discussed in section \ref{sec:shell}. The three hypothesis provide equally good fits to the \gr~data ($\Delta BIC < 2$) and lead to \gr~spectra consistent with the \gr~parameters given in section \ref{gnrl}.

\section{Conclusion}

\hess observations have led to the discovery of the distant, MC-interacting, SNR \snrG~in the VHE $\gamma$-ray domain. Although faint ($F(E>400\,\mathrm{GeV})\sim0.7\%$ of the Crab Nebula), its flux corresponds to a luminosity of  $\sim 10^{34}$~$\rm erg\,s^{-1}$, owing to its location in the Far 3 kpc Arm of the Galactic centre, at $\sim 11.5 \rm\, kpc$. The point-like shape of the VHE emission does not allow for an investigation of the morphology. Nonetheless, the combined \fermi~and \hess~spectrum, together with several other observational lines of evidence, strongly suggest that the \gr~emission results from the interaction between the SNR and the adjacent MC. By taking into account radio and X-ray data, the leptonically dominated scenarios for the origin of the \gr~emission are strongly disfavored, and $\pi^0$ decay from hadronic interactions requires a total energy content in CRs $n_{\rm H}W_{\rm p} \sim 5 \times 10^{51}$(d/11.5 kpc)$^{2}$ erg cm$^{-3}$. Although the \gr~spectrum and the inferred proton distributions are statistically compatible with a broken power-law and a power-law with exponential cutoff, the former shape is reminiscent of most of the $\gamma$-ray-emitting SNRs known to be interacting with MCs.

\begin{acknowledgements}
The support of the Namibian authorities and of the University of Namibia in facilitating the construction and operation of H.E.S.S. is gratefully acknowledged, as is the support by the German Ministry for Education and Research (BMBF), the Max Planck Society, the German Research Foundation (DFG), the French Ministry for Research, the CNRS-IN2P3 and the Astroparticle Interdisciplinary Programme of the CNRS, the U.K. Science and Technology Facilities Council (STFC), the IPNP of the Charles University, the Czech Science Foundation, the Polish Ministry of Science and Higher Education, the South African Department of Science and Technology and National Research Foundation, and by the University of Namibia. We appreciate the excellent work of the technical support staff in Berlin, Durham, Hamburg, Heidelberg, Palaiseau, Paris, Saclay, and in Namibia in the construction and operation of the equipment.
We thank the referee, Patrick Slane, for his comments and helpful suggestions.
\end{acknowledgements}

\bibliographystyle{aa}

\bibliography{G349}

\end{document}